\documentclass{amsart}

\RequirePackage{hyperref}
\usepackage{amsmath}
\usepackage{mathrsfs}
\usepackage{eufrak}
\usepackage[utf8]{inputenc} %unicode support
\usepackage[margin=3.6cm]{geometry}
\usepackage[linesnumbered,boxed,ruled,algonl]{algorithm2e}

\usepackage{graphicx} % LOCAL COPY BY NACHO

\usepackage{array,booktabs}
\usepackage{tikz}

\def\sC{{\mathscr C}}
\def\sD{{\mathscr D}}
\DeclareMathOperator*{\argmax}{arg\,max}

\begin{document}
\title{Router-level community structure of the Internet Autonomous Systems}
\author{Mariano~G.~Beiró$^{1,2}$ \and~Sebastián~P.~Grynberg$^{2}$ \and~J.~Ignacio~Alvarez-Hamelin$^{2,3,4}$}
\thanks{$^{1}$ISI Foundation, Via Alassio 11/c, 10126 Torino, Italy \\
\indent$^{2}$Fac. de Ingenier\'{\i}a (UBA), Paseo Col\'on 850, C1063ACV Buenos Aires, Argentina \\
\indent$^{3}$INTECIN (UBA-CONICET), Fac. de Ingenier\'{\i}a, Paseo Col\'on 850, C1063ACV Buenos Aires, Argentina \\
\indent$^{4}$ITBA, Av. Madero 399, C1106ACD Buenos Aires, Argentina}

%\author[a,b]{Mariano~G.~Beir\'{\o}$^{1,2}$\thanks{Corresponding author. E-mail: mariano.beiro@isi.it}}
%\author[b]{Sebasti\'{\a}n~P.~Grynberg$^{2}$}
%\author[b,c,d]{J.~Ignacio~Alvarez-Hamelin$^{2,3,4}$}
%\affil[a]{ISI Foundation, Via Alassio 11/c, 10126 Torino, Italy}
%\affil[b]{Facultad de Ingenier\'{\i}a, Universidad de Buenos Aires, Paseo Col\'on 850, C1063ACV Buenos Aires, Argentina}
%\affil[c]{INTECIN (UBA-CONICET), Facultad de Ingenier\'{\i}a, Universidad de Buenos Aires, Paseo Col\'on 850, C1063ACV Buenos Aires, Argentina}
%\affil[d]{ITBA, Av. Madero 399, C1106ACD Buenos Aires, Argentina}

\begin{abstract}

The Internet is composed of routing devices connected between them and organized into independent administrative entities: the Autonomous Systems. The existence of different types of Autonomous Systems (like large connectivity providers, Internet Service Providers or universities) together with geographical and economical constraints, turns the Internet into a complex modular and hierarchical network. This organization is reflected in many properties of the Internet topology, like its high degree of clustering and its robustness.

In this work, we study the modular structure of the Internet router-level graph in order to assess to what extent the Autonomous Systems satisfy some of the known notions of community structure. We show that the modular structure of the Internet is much richer than what can be captured by the current community detection methods, which are severely affected by resolution limits and by the heterogeneity of the Autonomous Systems. Here we overcome this issue by using a multiresolution detection algorithm combined with a small sample of nodes. We also discuss recent work on community structure in the light of our results.
%\noindent{\em {\bf Keywords:} Internet Topology, Community Structure, Autonomous Systems, Complex Networks}
\end{abstract}

\maketitle

\section{Introduction}

The Internet is a complex network composed of routing devices {\em (routers)} which are organized into administrative entities, the {\em Autonomous Systems (ASes)}. Each AS has its own routing policies and design criteria, which are based on technological, geographical and economical constraints and aim at maximizing the performance in terms of bandwidth, delay and resilience. ASes are nowadays identified with large carriers, ISPs, IXPs, CDNs, and universities, among others, and they are interconnected as the result of commercial agreements between them, either as peers or in provider-customer relationships.

ASes can also be classified into core ASes (once called Tier-1's), and peripheral ASes. The core ASes are typically the large carriers and some CDNs and are densely connected between them. Instead, most ISPs lay in the periphery and are connected to one or more core ASes. This core-periphery division provides the Internet with an interesting hierarchy. In fact, many large-scale properties as the resilience, scalability and small-worldness of the Internet arise largely as a consequence of its hierarchical structure and self-organization~\cite{caldarelli_lssadocn_2007, ravasz_hoicn_2003}. Understanding them might help developing better models of the Internet and improving its performance, by optimizing routing or reducing congestion, for example.

In this work we approach the hierarchical structure of the Internet at the router level by trying to identify the ASes as communities of nodes. We explore different notions of community structure, and we shall see that: {\em (a)} The goodness of the communities is in some way related to the core-periphery hierarchy of the ASes; {\em (b)} Though most of the ASes satisfy some notion of community, their largely variable sizes represent a hard problem for the community detection algorithms. We shall also propose a method for identifying ASes using samples of nodes.

Our work falls into the study of Internet topology. Some of the main results on this area can be found in~\cite{pastor2007evolution}. In particular, the study of community structure of the Internet at the Autonomous Systems level has been previously approached in~\cite{gregori1, gregori2, community_overlays, rossi2013topology} and, to the best of our knowledge, the community structure at the router level has only been addressed in~\cite{hirayama_2012} in the context of traffic queuing analysis. For the community detection problem in general, we address the reader to the survey by Fortunato~\cite{Santo201075}.

The paper is organized as follows: In the {\em Dataset} section we describe the Internet exploration that we used; then, in {\em Analysis methods and results} we apply several community detection algorithms to it, matching the obtained communities to the ASes; we also introduce a multiresolution community detection algorithm based on modularity maximization and we determine which is the best resolution for each AS; in the {\em Discussion} section we evaluate the results and explain their variability: why some ASes satisfy our notion of community while others do not. Finally, in the {\em Conclusions} we extract the main results of this work.

\section{Dataset}

The dataset used throughout this paper was provided by CAIDA\cite{CAIDA}. The CAIDA association performs daily explorations of the Internet by sending periodic IP packets ({\em probes}) towards random destinations from a set of sources called {\em monitors} (which nowadays count around $100$). These probes are sent with a {\tt traceroute} tool which exploits the ICMP functionality: by means of the ICMP protocol, the intermediate routers in a datagram transmission can provide the source with some information on the traversed path. Using this information, the CAIDA infrastructure builds a map of the Internet at the router level. Several problems must be resolved through this process, such as IP aliasing (the association of several IP addresses with one same router)~\cite{aliasres} and the existence of anonymous routers (routers which do not answer probes). Traceroute-based sampling is not a perfect sampling and has several biases~\cite{asta_traceroute}: some links are more frequently visited than others, and some links are probably not visited at all. This, together with the fact that the Internet is a fast evolving network, turns the map into a partial, approximate view of the Internet.

However, router-level maps of the Internet have several applications, like improving routing algorithms by taking into account the network topology, understanding information propagation, and studying the robustness and scale-free properties of networks~\cite{caldarelli_lssadocn_2007, pastor2004topology}. They can also be used for constructing maps of the Internet at the Autonomous Systems level~\cite{Chang01inferringas-level, Tangmunarunkit}.

\begin{figure}[!htb]
 \centering
 \includegraphics[width=12cm]{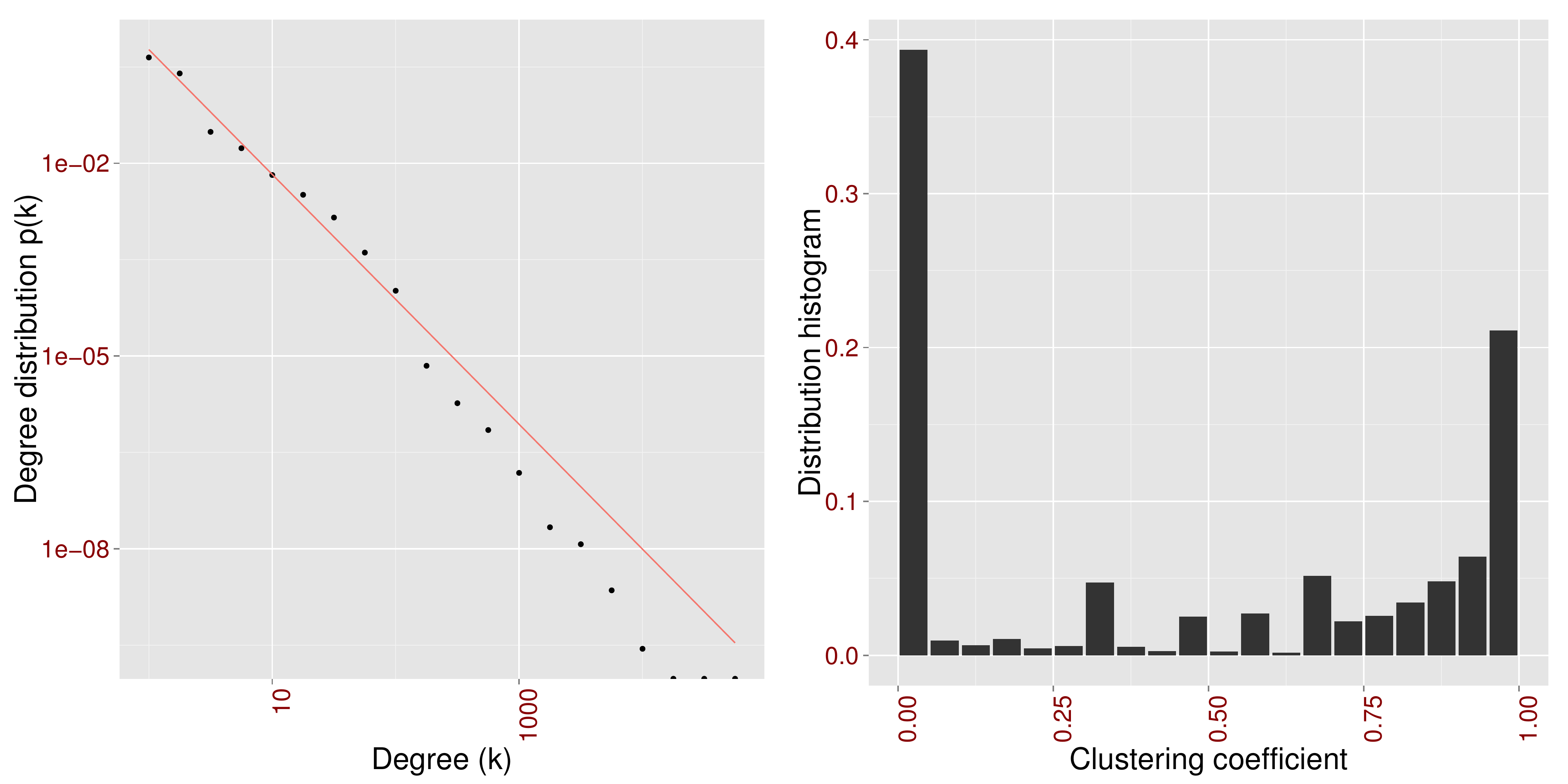}
  \caption{{Node degree and clustering coefficient.} In the left picture, the node degree distribution adjusted to a power law with $\alpha=1.94$ (max-likelihood estimation). In the right, a histogram of the clustering coefficient of the nodes (we omit nodes of degree $1$ in the computation).}\label{fig_degree_cc_dist_multi}
\end{figure}

\begin{figure}[!htb]
 \centering
 \includegraphics[width=12cm]{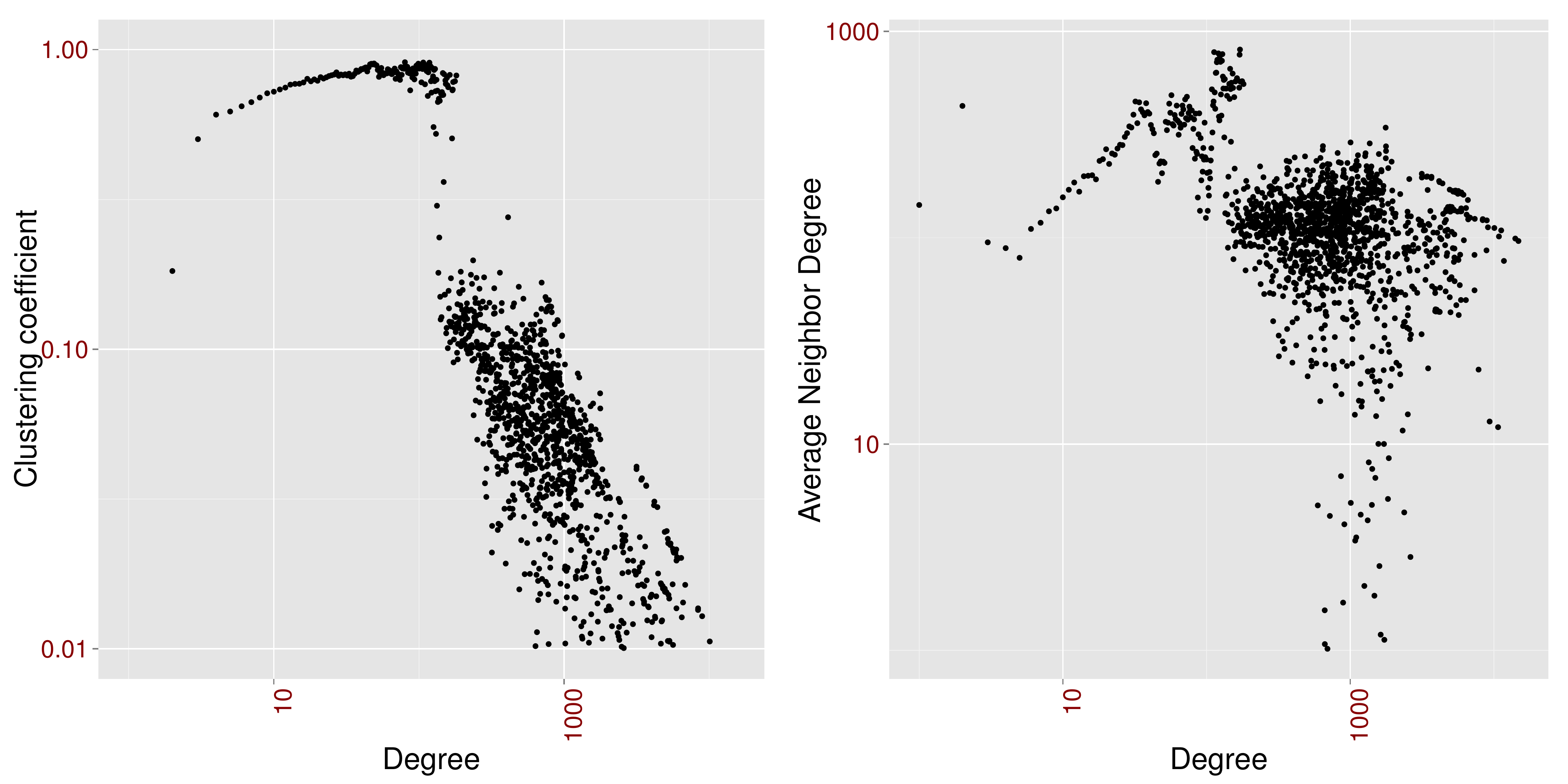}	
  \caption{{Average clustering coefficient and average neighbour degree as a function of node degree.} The left figure plots the clustering coefficient as a function of node degree (we omit nodes of degree $1$ in the computation). On the right figure, the average neighbour degree as a function of node degree.}\label{fig_cc_knn_multi}
\end{figure}

In this work we used the CAIDA router-level Internet map from October 19th, 2011~\cite{CAIDA2}. From the raw data we obtained an undirected graph consisting on $3,248,358$ nodes (the routers) connected by $14,083,946$ edges; $81.6\%$ of the nodes also contain information about their Autonomous System (AS) affiliation (for a description on how this information was inferred, consult~\cite{huffaker} and the CAIDA page~\cite{caida_data_kit}). The distributions for the degree and the clustering coefficient are shown in Figure~\ref{fig_degree_cc_dist_multi}. Figure~\ref{fig_cc_knn_multi} shows the average clustering coefficient and neighbour degree distribution as a function of the node degree.

\subsection{Graph preprocessing}

The graph contains some low-degree structures which make it difficult to analyse its community structure without some preprocessing. Regarding the AS affiliation, for example, the dataset contains $21,108$ different ASes, of which $8,023$ have less than $5$ nodes, and $865$ ASes contain $1$ single node of degree $1$. We also detected the presence of many {\em tendrils} (sequences of degree-$2$ nodes finishing with a degree-$1$ node) involving $1,393,846$ nodes, which do not help to detect community structure. We removed these nodes by computing the $2$-core of the graph~\cite{batagelj_new}.

We also observed many sequences of degree-$2$ nodes finishing with higher degree nodes in their endpoints; we call these sequences {\em chains}. We found a negative correlation of $-0.7$ between the events {\em having degree-$2$} and {\em not having AS affiliation}, as shown in Table~\ref{table_correlation_events}; in fact, $88.9\%$ of nodes without AS affiliation belong to chains, while $97.2\%$ of the remaining nodes do contain AS affiliation. Some of the chains are due to the presence of {\em MPLS tunnels}~\cite{pansiot} or other types of tunnels: $97.5\%$ of nodes involved in chains and without AS affiliation have private IP addresses (i.e. their interfaces did not answer the traceroute probes). We could classify $64.7\%$ of the chains as internal to the ASes, and $33.8\%$ of them as inter-AS chains.

\begin{table}[!htb]
\caption{\label{table_correlation_events}{\bf Correlation between having degree $2$ and having AS affiliation.} The table shows the joint distribution of the indicator variables $degree2$:{\em having degree $2$} and $AS$:{\em having AS affiliation}.}
	\centering
      \begin{tabular}{ccc}
        \hline
        $degree2$ $\setminus$ $AS$   & True  & False\\ \hline
        True & 0.12 & 0.16 \\
        False & 0.70 & 0.02 \\ \hline
      \end{tabular}
\end{table}

\tikzstyle{vertexN}=[circle,fill=black!80,draw,minimum size=10pt,inner sep=0pt]
\tikzstyle{vertexB}=[circle,fill=white!20,draw,minimum size=10pt,inner sep=0pt]
\tikzstyle{vertexG}=[circle,fill=gray!80,draw,minimum size=10pt,inner sep=0pt]

Thus, we decided to clear the graph by restricting the analysis to the $2$-core and replacing all the chains by an edge between their endpoints. We found $531,625$ chains of different types, as resumed in Table~\ref{table_chains}. The resulting graph after this process has $1,119,672$ nodes and $11,742,947$ edges. $99.2\%$ of the nodes have AS affiliation now, and just $15,288$ different ASes are found.

\begin{table}[!htb]
\caption{\label{table_chains}{\bf Most common cases of chains of routers.} Black and gray colours represent some hypothetical ASes; white nodes do not have AS affiliation. Many of these chains are caused by tunnels (like MPLS tunnels); others are probably due to a tie in the AS affiliation algorithm. After removing these chains, $99.2\%$ of the nodes have AS affiliation.}
	 \centering
	\renewcommand{\arraystretch}{2.0}
      \begin{tabular}{>{\centering\arraybackslash}m{1.6in} >{\centering\arraybackslash}m{1.0in} >{\centering\arraybackslash}m{1.4in}}
        \hline
	Chain & Number of cases & Typical configuration\\
	\hline \hline
		\begin{tikzpicture}[shorten >=1pt,-]
		  \foreach \name/\x in {1/1, 2/2, 3/3}
		    \node[vertexN] (G-\name) at (\x,0) {};
		  \foreach \from/\to in {1/2,2/3}
		    \draw (G-\from) -- (G-\to);
		\end{tikzpicture} & 193398 & Internal AS connection\\
		\begin{tikzpicture}[shorten >=1pt,-]
		  \foreach \name/\x in {1/1, 3/3}
		    \node[vertexN] (G-\name) at (\x,0) {};
		  \foreach \name/\x in {2/2}
		    \node[vertexB] (G-\name) at (\x,0) {};
		  \foreach \from/\to in {1/2,2/3}
		    \draw (G-\from) -- (G-\to);
		\end{tikzpicture} &  90307 & Internal AS tunnel\\
		\begin{tikzpicture}[shorten >=1pt,-]
		  \foreach \name/\x in {1/1}
		    \node[vertexN] (G-\name) at (\x,0) {};
		  \foreach \name/\x in {2/2}
		    \node[vertexB] (G-\name) at (\x,0) {};
		  \foreach \name/\x in {3/3}
		    \node[vertexG] (G-\name) at (\x,0) {};
		  \foreach \from/\to in {1/2,2/3}
		    \draw (G-\from) -- (G-\to);
		\end{tikzpicture} &  86775 & Inter-AS tunnel\\
		\begin{tikzpicture}[shorten >=1pt,-]
		  \foreach \name/\x in {1/1}
		    \node[vertexN] (G-\name) at (\x,0) {};
		  \foreach \name/\x in {2/2, 3/3}
		    \node[vertexB] (G-\name) at (\x,0) {};
		  \foreach \name/\x in {4/4}
		    \node[vertexG] (G-\name) at (\x,0) {};
		  \foreach \from/\to in {1/2,2/3, 3/4}
		    \draw (G-\from) -- (G-\to);
		\end{tikzpicture} &  57988 & Inter-AS tunnel\\
		\begin{tikzpicture}[shorten >=1pt,-]
		  \foreach \name/\x in {1/1}
		    \node[vertexN] (G-\name) at (\x,0) {};
		  \foreach \name/\x in {2/2, 3/3}
		    \node[vertexB] (G-\name) at (\x,0) {};
		  \foreach \name/\x in {4/4}
		    \node[vertexN] (G-\name) at (\x,0) {};
		  \foreach \from/\to in {1/2,2/3, 3/4}
		    \draw (G-\from) -- (G-\to);
		\end{tikzpicture} &  32473 & Internal AS tunnel\\
		\begin{tikzpicture}[shorten >=1pt,-]
		  \foreach \name/\x in {1/1}
		    \node[vertexN] (G-\name) at (\x,0) {};
		  \foreach \name/\x in {2/2, 3/3, 4/4}
		    \node[vertexB] (G-\name) at (\x,0) {};
		  \foreach \name/\x in {5/5}
		    \node[vertexG] (G-\name) at (\x,0) {};
		  \foreach \from/\to in {1/2,2/3, 3/4, 4/5}
		    \draw (G-\from) -- (G-\to);
		\end{tikzpicture} &  28774 & Inter-AS tunnel \\	\hline
      \end{tabular}
\end{table}

\section{Analysis methods and results}

We analysed the community structure of the Internet router-level graph by applying several community detection methods:
\begin{itemize}
\item Infomap, by Rosvall and Bergstrom, based in the description length~\cite{rosvall_aitffrcsicn_2007}.
\item LPM, the Label Propagation Method by Raghavan {\em et al.}~\cite{raghavan_nltatdcsilsn_2007}, which performs a diffusion process on the graph.
\item Deltacom, an efficient greedy algorithm for modularity optimization introduced here.
\item Louvain, a fast modularity optimization algorithm~\cite{blondel_fuociln_2008}.
\item CommUGP, a local community detection method~\cite{ocwafgp_beiro_2013}.
\end{itemize}

In the following subsection we shall introduce Deltacom, an algorithm for community detection based on modularity maximization. Later on, we shall evaluate the results using a similarity metric.

\subsection{A multiresolution community extraction method}

Deltacom is based on the optimization of modularity. We recall that modularity was introduced by Newman in~\cite{newman_faecsin_2004, clauset_fcsivln_2004} and, since then, several methods for its maximization have been proposed, like the one by Guimerà {\em et al.} based on simulated annealing~\cite{guimera_cocnmaur_2005}, the extremal optimization method by Duch and Arenas~\cite{duch_cdicnueo_2005}, the fast greedy algorithm by Blondel {\em et al.}~\cite{blondel_fuociln_2008}, or the multilevel algorithm by Noack and Rotta~\cite{noack_mlafmc_2009}.

The following two results about modularity are fundamental to our present work: in~\cite{fortunato_rlicd_2007} Fortunato and Barthelemy showed that modularity has a resolution limit, i.e., its maximization tends to put small communities together when they are connected among them; in~\cite{reichardt_smocd_2006} Reichardt and Bornholdt observed that modularity can be understood as the Hamiltonian of a ferromagnetic Potts model. Thus, maximizing the modularity implies finding the ground-state of this model, and the authors developed a simulated annealing based procedure for doing it.

Deltacom considers the optimization of the modularity $Q$ as a particular case of the optimization of a more general functional $Q_t$, with a resolution parameter $t$, in which modularity corresponds to a normalized resolution, i.e., $t=1$. The source of our idea can be traced back to the $\gamma$ resolution parameter in~\cite{reichardt_smocd_2006}, and has also been followed by Lancichinetti and Fortunato in~\cite{Lancichinetti2011}. In~\cite{reichardt_smocd_2006}, the maximization is based on simulated annealing, it must run at one single resolution each time, and its computational complexity is very high. The advantage of our method is that the resolution evolves dynamically, so that all the partitions at different resolutions can be produced during one single run, and with a low computational complexity.

Deltacom is based on an agglomerative greedy algorithm; each step of the agglomerative process is a local maximum for $Q_t$, i.e., the generalized modularity, at some particular resolution $t$. In other words, for each $t$-value a community partition is found, which is locally $t$-optimal in some sense. As these communities are joined, the resolution shrinks and $t$ gets smaller. For $t=1$, the structure is a local maximum for the classical modularity. In other words, we observe the graph as by using a magnifying glass, obtaining partitions at every resolution level. In particular, partitions at higher resolution levels are always refinements (in a mathematical sense) of those at lower ones.

Here we present a brief description of how the algorithm works. Further mathematical details on the properties of $Q_t$ and of our maximization algorithm can be found in~\cite{alvarezh_owopimn_2010}. 

\subsubsection*{Newman's modularity}

Given a graph $G=(V,E)$ and a partition of it, $\sC$, the modularity of the partition is defined as~\cite{clauset_fcsivln_2004}:
\begin{equation}
\label{eqnewman}
Q(\sC) = \frac{1}{2m}\sum_{(i,j)\in V}{\left[A_{ij}-\frac{k_i k_j}{2m}\right]}\delta(i, j)\enspace,
\end{equation}

\noindent where $m$ is the amount of edges, $A=(A_{ij}), i,j\in V$ is the adjacency matrix of the graph, $k_i$ is the degree of the node $i$ and $\delta(i,j)$ is the indicator function that points out if nodes $i$ and $j$ belong to the same community in the partition $\sC$.

The expression on the right side of Eq.~\ref{eqnewman} compares the internal connections in the communities (represented by the $A_{ij}\cdot\delta(i,j)$ product) with the expected number of connections under a random graph with the same expected degrees in the nodes and the same communities. It can be restated as a sum throughout the communities, in the following way:
\begin{equation}
Q(\sC) = \frac{1}{2m}\sum_{C\in \sC}{\left[e_C-\frac{k_C^2}{2m}\right]}\enspace,
\end{equation}

\noindent where $k_C$ is the sum of degrees of the nodes in each community $C$, and $e_C$ is twice the number of internal edges of it.

Now, in many agglomerative clustering methods, communities are joined by pairs at each step, until finding a local maximum. Joining two communities $C$ and $C'$ during the process produces the following variation in the $Q$ functional:
\begin{equation}
\Delta Q = \frac{1}{2m}\left(e(C,C')-\frac{k_C\cdot k_{C'}}{2m}\right)\enspace,
\end{equation}

\noindent where $e(C,C')$ is twice the number of edges between $C$ and $C'$.

\subsubsection*{Introducing a resolution parameter}

We introduce a resolution parameter $t$ as
\begin{equation}
Q_t(\sC) = \frac{1}{2m}\sum_{C\in \sC}{\left[e_C-t\cdot\frac{k_C^2}{2m}\right]}\enspace.
\end{equation}

Now the functional variation after joining two communities $C$ and $C'$ becomes:
\begin{equation}
\Delta Q_t = \frac{1}{2m}\left(e(C,C')-t\cdot\frac{k_C\cdot k_{C'}}{2m}\right)\enspace.
\end{equation}

It is clear that a positive $\Delta Q_t$ value for a particular $t$ also implies a positive (and even higher) $\Delta Q_{t'}$ for any resolution $t'<t$. In other words, any agglomerative process which monotonically increases $\Delta Q_t$ also serves as a process monotonically increasing $\Delta Q_{t'}$ for any $t'<t$.

It is also immediate that a very large $t$ value discourages any join, because $\Delta Q_t$ is negative for every pair of communities. It can also be shown that for $t$ larger enough the global maximum for $Q_t$ would have each node isolated in its own community.

Thus, what we propose is to start with a large enough $t$ (so that the optimal initial partition $\sC_t$ has as many communities as nodes) and then to decrease $t$ as little as possible so that a join can be made, i.e., so that some $\Delta Q_t$ becomes positive. This $t$ value is just
\begin{equation}
t(\sC) = \max_{C,C'\in\sC}\left\{\frac{e(C,C')\cdot2m}{k_C\cdot k_{C'}}\right\}\enspace.
\end{equation}

When a local maximum is found for some $t$ (i.e., no pair of communities can be joined without decreasing $\Delta Q_t$), we decrease the resolution to $t'$ such that $\Delta Q_t'$ becomes positive for some pair of communities, following the previous formula. The agglomerative process continues until obtaining as many communities as connected components of the graph, and the obtained result is a set of locally optimal partitions $\sC_t$ for every resolution $t$, and such that the finer partitions are refinements of the coarser-grained ones. Even more, these partitions are what we call {\em weakly optimal} partitions, in the sense that not only the join of any pair of communities $C,C'$ would decrease $Q_t$, but also any coarser partition (i.e. obtained by joining its communities in any way) would also decrease it.

The code for the Deltacom algorithm is freely available at SourceForge~\cite{sourceforge}; given a graph as a list of edges, it produces a set of community partitions for every resolution value.

The formalization of the process is contained in Table~\ref{table_alg}. For a complexity analysis, we shall consider keeping a set of lists $L_C$, one for each community, containing its neighbour communities $C'$ and the edge cut with each of them. The lists shall be ordered by a community identifier. The initial construction of these structures has time complexity $\mathcal{O}(m\cdot log(m))$ (as each list has size $s_C$, with $\sum_{C}{s_C}=m$, and $\sum_{C}{s_C\cdot log(s_C)}\leq m\cdot log(m)$). The {\tt while} loop starting at line {\tt 1.6} has at most $n$ cycles; each of these cycles has three steps: {\em (a)} finding a pair of communities with $\Delta(Q_t)=0$; {\em (b)} joining the communities and; {\em (c)} decreasing $t$. The first step implies traversing all the lists; each $\Delta(Q_t)$ computation is $\mathcal{O}(1)$ so this step's complexity can be bounded to $\mathcal{O}(m)$. Joining two communities implies merging their lists into one, and also updating the lists of their neighbour communities. As all the lists are ordered, this can be done by traversing those lists, which we can bound to $\mathcal{O}(m)$. Finally, decreasing $t$ implies traversing all the lists and computing $\frac{e(C,C')\cdot2m}{k_C\cdot k_{C'}}$ for each element. This is again $\mathcal{O}(m)$. Putting all together, we have at most $n$ cycles with $\mathcal{O}(m)$. The final complexity is then $\mathcal{O}(nm)$.

While $\mathcal{O}(nm)$ is a theoretical upper bound for a general graph, the practical running time for sparse graphs can be greatly reduced with some considerations: step {\em (c)} gives all the information for the next step {\em (a)}, because the community pair which maximized $t'$ in line {\tt 1.11} is the same that will have $\Delta Q_t=0$ in the next iteration; also, by keeping a search structure like an ordered tree with all the pair of connected communities $C,C'$ ordered by $\frac{e(C,C')\cdot2m}{k_C\cdot k_{C'}}$ decreasingly, there is no need to traverse all the lists in each iteration, but we must just update the modified values into this tree and choose the community pair at its head as the one maximizing $t'$. For sparse graphs, this process has an upper complexity bound of $\mathcal{O}(m\cdot log(m)\cdot s_C^{max})$, where $s_C^{max}$ is the maximum number of neighbour communities that a community may have at any time of the process.

\begin{table}[!htb]
\caption{\label{table_alg}{\bf Pseudo-code for the multiresolution modularity optimization with Deltacom.} This algorithm produces weakly optimal partitions for the generalized modularity $Q_t$. In particular, $\sC_1$ is a weakly optimal partition for the classical modularity $Q$. The notation $\sC_{(a,b]}$ represents that the partition is weakly optimal for every resolution in the $(a,b]$ interval.}
\begin{algorithm}[H]
  \caption{Deltacom: Multiresolution modularity optimization}\label{algorithm1}
  \SetLine
  \KwIn{A graph $G=(V,E)$}
  \KwOut{A set of weakly optimal partitions $\sC_t$ for every resolution value $t_{\min}\leq t\leq t_{\max}$, such that $\sC_{t'}\prec\sC_t$ whenever $t'<t$}
  \Begin{
     $\sD = \{ \{ v \}, v\in V \}$ {\tt \#Initial partition} \\
     $t_{\max} = \max_{C,C'\in\sD}\left\{\frac{e(C,C')\cdot2m}{k_C\cdot k_{C'}}\right\}$ {\tt \#Initial t} \\
     $t = t_{\max}$ \\
     $\sC_t=\sD$ \\
     \While {$|\sD|>1$} {
         \While {there exists a pair $(C, C')$ in $\sD$ such that $\Delta Q_t(C,C')=0$}{
	    $\sD=\sD\setminus\{C,C'\}\cup\{C\cup C'\}$
         }
	 \uIf {$|\sD|>1$} {
           $t' = \max_{C,C'\in\sD}\left\{\frac{e(C,C')\cdot2m}{k_C\cdot k_{C'}}\right\}$ {\tt \#Decreasing t} \\
           $\sC_{(t',t]}=\sD$ {\tt \#Output partition for the range $(t', t]$ is finished} \\
         } \Else {
	   $t' = t$  {\tt \#In the last step we cannot decrease t} \\
           $\sC_t=\sD$ {\tt \#Last partition; composed of one large community} \\
  	 }
   	 $t = t'$
     }
     $t_{\min}=t$ {\tt \#Minimum resolution; all the graph into one large community} \\
   }
\end{algorithm}
\end{table}

\subsection{Finding ASes through similarity maximization}

We applied the five forementioned algorithms (Infomap, LPM, Deltacom at $t=1$, Louvain and CommUGP) to the CAIDA dataset. Each of these methods produces a partition $\sC$ of the graph. In order to determine if the communities in these partitions are related to the ASes, we shall use this criterion: for each AS in the graph, and given a partition $\sC$, we find the community $C\in\sC$ which maximizes the {\em Jaccard similarity}:
\[
J(C, AS) = \frac{|C \cap AS|}{|C \cup AS|}\enspace.
\]

This most similar community shall be called $C_{AS}^{(1)}$:
\[
C_{AS}^{(1)} = \argmax_{C\in\sC} J(C, AS)\enspace.
\]

And this maximum Jaccard similarity value will be called the {\em recall score} of the AS (according to~\cite{PhysRevE.90.062805}) and noted as $R_1(AS)$:
\begin{equation}
\label{eq_j1}
R_1(AS) = J(C_{AS}^{(1)}, AS)\enspace.
\end{equation}

\begin{figure}[!htb]
 \centering
 \includegraphics[width=12cm]{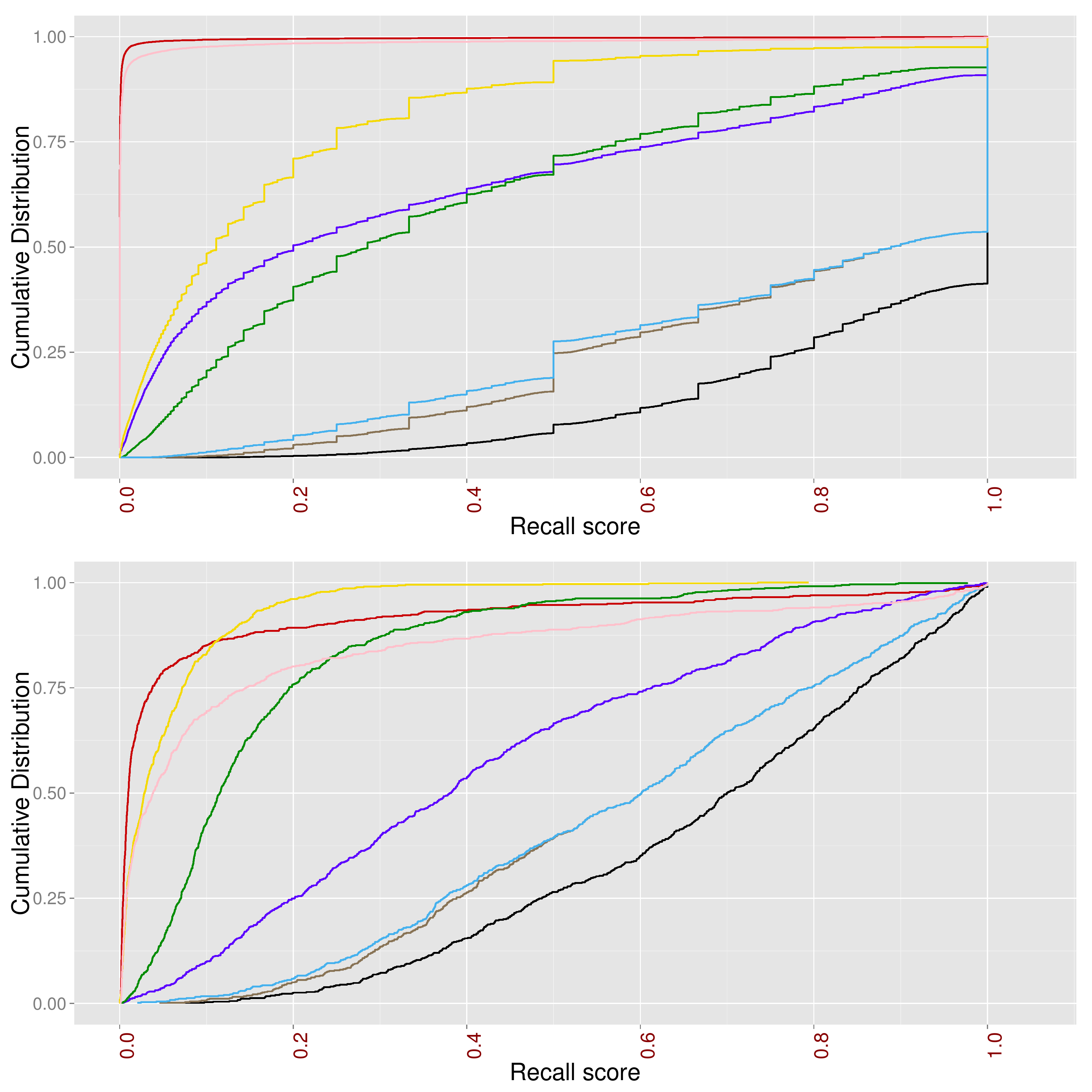}
  \caption{{Cumulative distribution of the AS recall score.} {\em (Upper)} The curves represent the cumulative distribution of the recall scores over all the ASes for different community discovery methods: $R_1(AS)$ is shown for Deltacom at $t=1$ (i.e., modularity maximization) {\em (red)}, LPM {\em (yellow)}, Louvain modularity maximization {\em (pink)}, CommUGP {\em (green)} and Infomap {\em (blue)}. If we we find the best resolution $t$ for each AS using the Multiresolution Deltacom algorithm, we get the {\em black} curve, corresponding to $R_2(AS)$. The {\em brown} curve represents the Multiresolution Deltacom using the resolution value $\tilde{t}$ predicted by the regression line for each AS, and then computing $R_1(AS)$ in $\sC_{\tilde{t}}$. Finally, if we use $15\%$ of the AS routers when matching the partition at $\sC_{\tilde{t}}$ for each AS, then we get the {\em light blue} curve. {\em (Lower)} The same methods as in the upper picture, but restricted to ASes containing at least $100$ nodes.}\label{results_jaccard_prediction}
\end{figure}

In this way, we can evaluate each of the methods by observing the empirical cumulative distributions of the recall score of the ASes, $R_1(AS)$ (Eq.~\ref{eq_j1}). Those methods with left-skewed cumulative distributions (i.e., recall values closer to $1.00$) tend to better reproduce the Autonomous System structure. The cumulative distributions for the different methods are part of Figure~\ref{results_jaccard_prediction}: {\em blue} for Infomap, {\em yellow} for LPM, {\em red} for Deltacom, {\em green} for CommUGP and {\em pink} for Louvain. Small ASes with just a few nodes may introduce some noise into the comparison (firstly, because ASes with two or three nodes are not relevant for the analysis and secondly, because due to exploration biases sometimes their nodes are not even connected) so we also plot the cumulative distribution restricted to ASes of at least $100$ nodes.

We can clearly see that the five methods fail to reproduce the ASes structure. In the best performing one, Infomap, only $50\%$ of the large ASes reach a recall of $0.4$, which is however small. A premature conclusion might state that the Internet graph does not have communities at the router level, or at least that they are not related with the Autonomous Systems. But we will show that this is not the case, and that the failure of the methods is in part due to the largely variable sizes, internal structures and functions of the Internet ASes. However, the structure of the Internet graph at the router level can be better captured by using multiresolution methods.

In order to assess our hypothesis we shall exploit the resolution parameter $t$ of Deltacom, which provides us with a new level of freedom for matching the ASes in the community structure. For each $t$ value we have a partition $\sC_t$. Thus, in this case we define the most similar community for an Autonomous System, $C_{AS}^{(2)}$, as:
\begin{align}
C_{AS}^t =& \argmax_{C\in\sC_t}{J(C, AS)}\enspace,\; t\in[t_{\min}, t_{\max}]\enspace, \notag \\
C_{AS}^{(2)} =& \argmax_{\{C_{AS}^t: \; t\in[t_{\min}, t_{\max}] \}}{J(C_{AS}^t, AS)}\enspace, \notag
\end{align}

and the recall score as:
\begin{equation}
\label{eq_j2}
R_2(AS) = J(C_{AS}^{(2)}, AS)\enspace.
\end{equation}

In this way, the Jaccard similarity allows us to capture the moment in which each AS is formed. With this adjustments, the results clearly improve. The recall score $R_2(AS)$ (Eq.~\ref{eq_j2}) over all the ASes is $0.87$ (and $0.67$ over the ASes of at least $100$ nodes). % and the weighted average is $0.69$.
The new cumulative distribution is presented as the black curve in Figure~\ref{results_jaccard_prediction}, and we shall call it {\em Deltacom Multiresolution}, as we are exploring all the communities at all possible resolutions.

\begin{figure}[!htb]
 \centering
 \includegraphics[width=12cm]{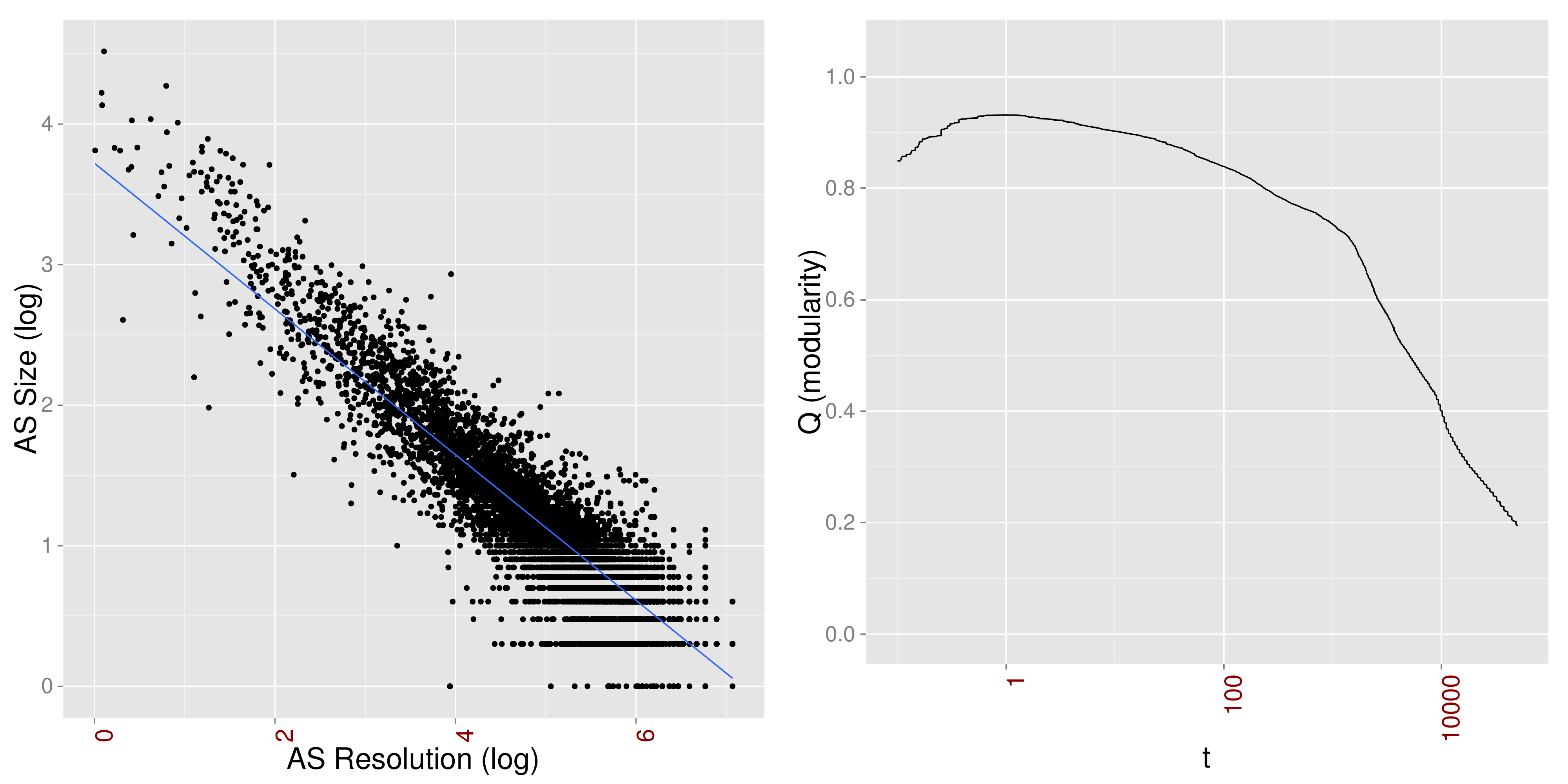}
  \caption{{Correlation between the AS size and resolution, and evolution of $Q$ in Deltacom.} The left figure plots the correlation between the AS size (in terms of the amount of routers) and the resolution at which it is best detected. The correlation coefficient between size and resolution in the log-scale is $-0.93$. A linear regression model between them gives the line $y=4.09-0.60x$ shown in blue, with a determination of $r^2=0.87$. On the right figure, the evolution of classical modularity ($Q_1$) as a function of $t$ during the agglomerative process of Deltacom.}\label{corr_t_size}
\end{figure}

The observed improvements tell us that each Autonomous System has a {\em natural} resolution at which it can be best observed. There is an important correlation between this resolution and the size of the AS, as the left picture in Figure~\ref{corr_t_size} shows. The right picture in Figure~\ref{corr_t_size} shows the evolution of classical modularity $Q$ for the Internet graph as a function of the resolution $t$, i.e. as the algorithm evolves (from right to left). As one of the properties of our algorithm, $Q_t$ has its maximum at $t$ for every possible $t$, so that the classical $Q$ is maximal at $t=1$.

\subsection{Detecting ASes from samples}

The previous results show us that different ASes have different resolutions, and that we cannot expect to get all the ASes with classical community detection methods or at some particular resolution value. However, our results with the multiresolution method are ideal, in the sense that we used the information of the AS structure in order to know at which resolution to stop for each AS. Now we wonder whether we might retrieve the ASes using a minimal amount of information: their size, and a small sample of their nodes. We observed that there is a strong dependence between the AS size and the AS resolution, so that we shall use the linear regression in Figure~\ref{corr_t_size} to approximate the best resolution for each AS, $\tilde{t}$. If we look for the most similar community to each AS restricted to the partition produced by Deltacom at that approximate resolution, $\sC_{\tilde{t}}$, then we obtain the brown curve in Figure~\ref{results_jaccard_prediction}. %The comparison between the brown and black curves gives an idea of how precise the regression between size and resolution is.

Lastly, we shall consider that we only know $15\%$ of the nodes at random. Suppose that $AS_{sample} \subset AS$ is the known subset for each $AS$. We shall estimate the most similar community using the $AS_{sample}$ and the partition at resolution $\tilde{t}$:
\begin{align}
\hat{C}_{AS}^{(2)} =& \argmax_{C\in\sC_{\tilde{t}}} J(C, AS_{sample})\enspace, \notag\\
R_3(AS) =& \hat{R}_2(AS) = J(\hat{C}_{AS}^{(2)}, AS)\enspace, \label{eq_j3}
\end{align}

\noindent i.e., we only consider the nodes of the AS which are part of our sample when choosing the most similar community. Instead, for computing the recall score $R_3(AS)$ (Eq.~\ref{eq_j3}) we consider all the nodes of the AS, so that the method can be compared against the others. The results for $R_3(AS)$ are those of the light blue curve in Figure~\ref{results_jaccard_prediction}.

The closeness between the brown and light blue curves proves that we do not need to know the whole AS in order to identify it as a community at a certain step of the modularity optimization process: $15\%$ of the nodes is enough in order to get as good a result as if we knew all the AS nodes. However, the distance between the brown and black curve tells us that the linear regression between size and resolution is just a rough approximation. In conclusion, the sample-based method seems to be very successful, and is only limited by the error of the linear regression. We can thus identify many of the ASes just knowing their size and a small sample of their nodes.

\section{Discussion}

The results in Figure~\ref{results_jaccard_prediction} show a type of stochastical dominance among the methods. The lowest curves are the ones most left-skewed and thus represent better correspondence between the communities identified by maximizing the Jaccard similarity and the ASes. In particular, the area above each curve is an interesting measure of the goodness of the method, and it also represents the mean recall score over all the ASes. The list of areas is presented in Table~\ref{means_methods}, computed for the cumulative distributions of ASes with at least $100$ nodes.

\begin{table}[!htb]
\caption{\label{means_methods}{\bf Average recall scores of the methods}. Average recall score (or area above the cumulative distribution curve) for the different methods, over the ASes with at least $100$ nodes.}
      \centering
      \begin{tabular}{lccc}
        \hline
        Method 					& avg $R(AS)$ \\ \hline
        LPM					& 0.05 \\
        Deltacom (t=1)				& 0.08 \\
        Louvain 				& 0.13 \\
        CommUGP 				& 0.16 \\
        Infomap 				& 0.41 \\
        {\bf Deltacom Multiresolution+Regression line+AS samples} 	& 0.58 \\
        Deltacom Multiresolution+Regression line 		& 0.59 \\
        Deltacom Multiresolution (Best Resolution) 		& 0.67 \\ \hline
      \end{tabular}
\end{table}

\begin{figure}[!htb]
 \centering
 \includegraphics[width=12cm]{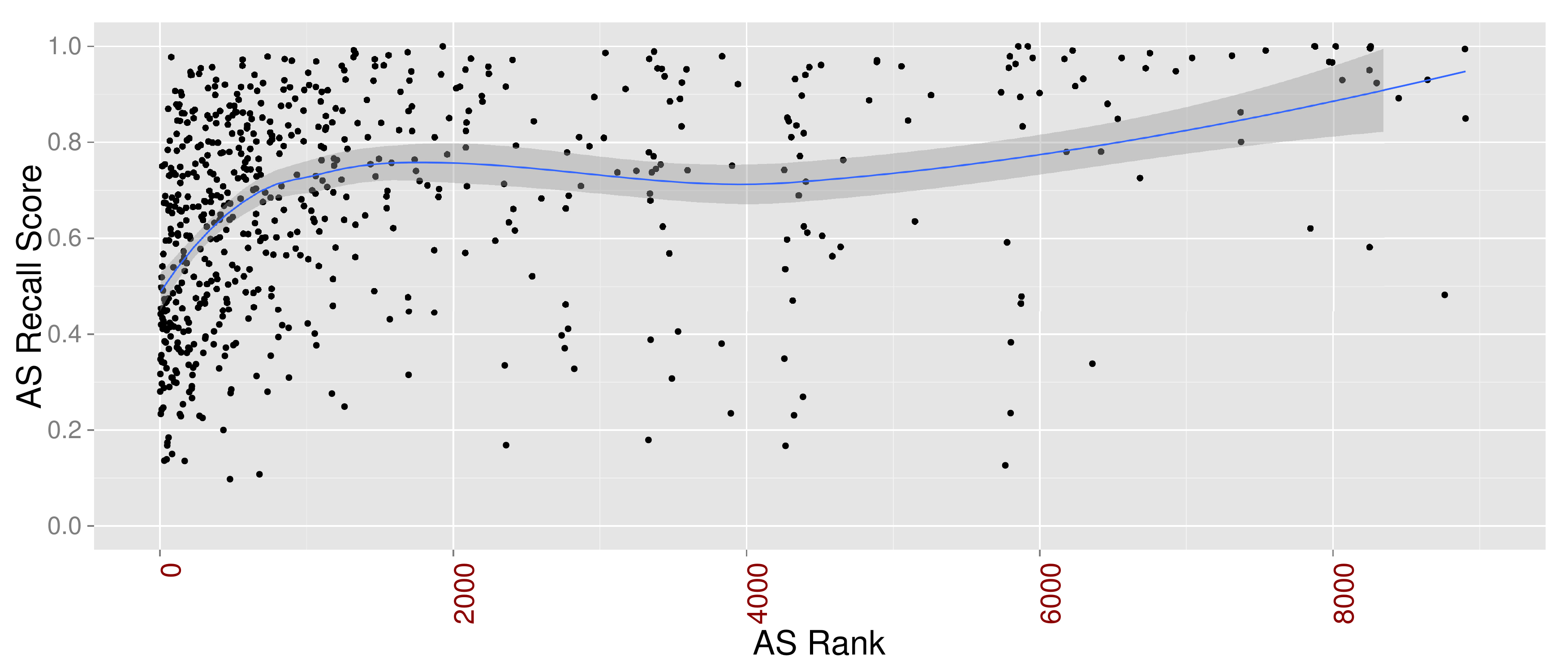}
  \caption{{AS Rank vs. AS Recall score.} Each point in this figure represents an Autonomous System, with the x-coordinate being its {\em AS rank} according to CAIDA~\cite{CAIDA} and its y-coordinate the recall score obtained with the Multiresolution Deltacom algorithm, $R_2(AS)$ (Eq.~\ref{eq_j2}). Only the ASes with at least $100$ nodes are represented here. Most of the low recall scores (less than $0.40$) correspond to high-ranked ASes, which are usually Tier-1's. The continuous curve was obtained with the LOESS local regressor, and the shadowed area represents the $95\%$ confidence interval for the curve.}\label{fig_as_rank}
\end{figure}

The difficulties of classical modularity maximization due to the resolution limit and the degeneracy of its maxima are manifested in the poor average values obtained by Deltacom at $t=1$ ($0.08$) and by Louvain ($0.13$) for the recall score of ASes with at least $100$ nodes. But Deltacom Multiresolution has a good performance when we explore all the possible resolutions, showing that most of the Autonomous Systems do have a community structure ($80\%$ of them have a recall over $0.70$). Instead, most of the ASes which were not detected by Deltacom Multiresolution belong to the core of the Internet: in Figure~\ref{fig_as_rank} we plot a regression curve of the recall score on the {\em AS rank}. The AS rank is a ranking of ASes designed by CAIDA and based on the customer cone size (the set of ASes that an AS can reach using customer links~\cite{2013relationships}). We observe that most of the ASes found with a low recall have a high AS rank and belong to large carriers. Regarding the scarce points with low AS Rank and low recall, we observed that some of them belong to IXPs and CDNs, which also belong to the Internet backbone but have a small customer cone.

\begin{figure}[!htb]
 \centering
\includegraphics[width=11cm]{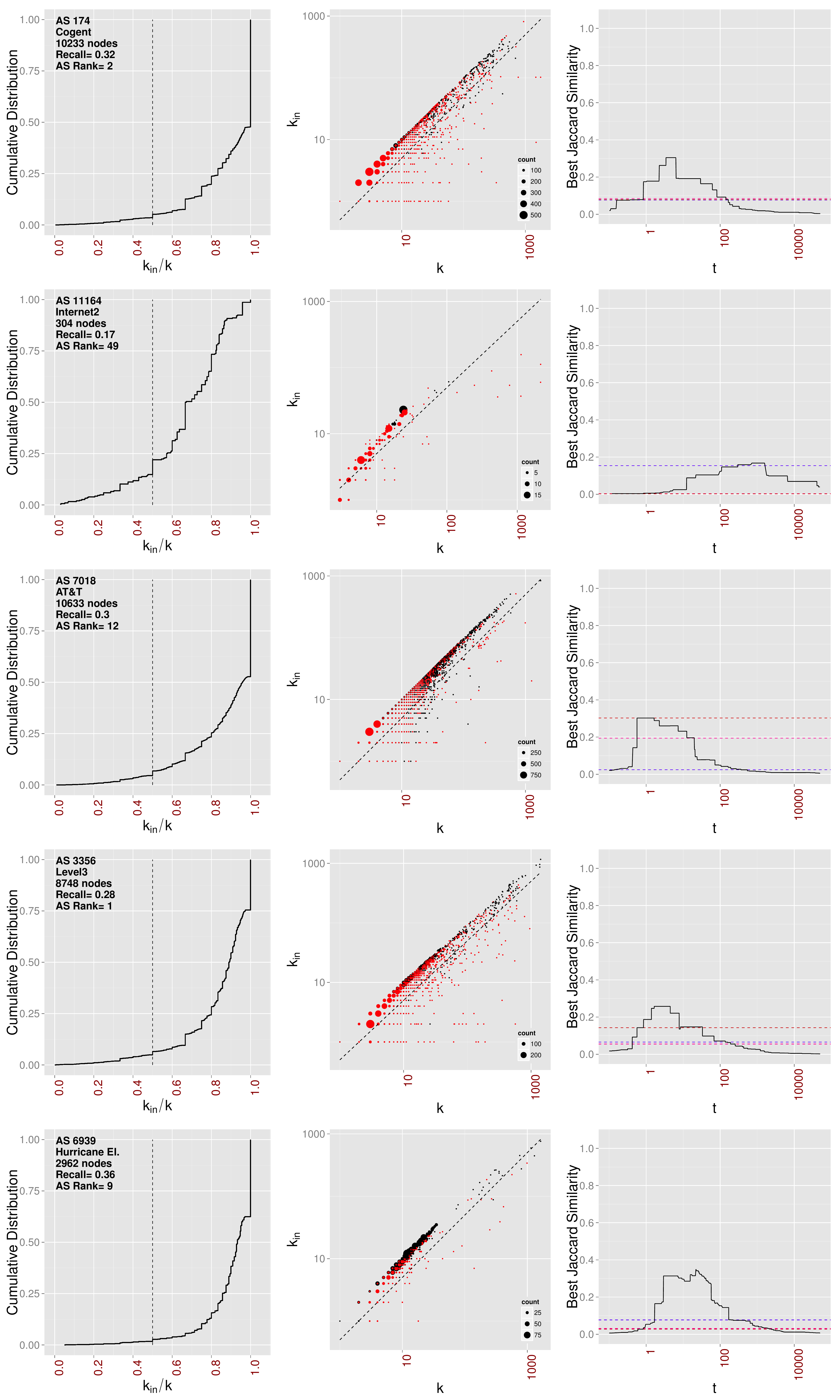}
  \caption{{High-ranked ASes with bad recall score.} Each row represents an AS. The left column plots the distribution of the internal node connections, $k_{in}$, normalized by the node degree $k$. The vertical dashed line at $k_{in}/k=0.5$ represents the notion of strong community. The central column plots the internal node degree, $k_{in}$, vs. the node degree, $k$. The point size represents the number of nodes with that $(k_{in}, k)$ pair; the diagonal line represents the condition $k_{in}/k=0.5$. The node colour indicates if the node belongs to the most similar community retrieved by Multiresolution Deltacom (black) or not (red). The right column plots the best Jaccard similarity obtained with Multiresolution Deltacom at resolution $t$ (its maximum being the recall). The horizontal dashed lines represent the recall score for Deltacom at $t=1$ (classical modularity) in red, for Infomap in blue and for Louvain in pink.}\label{ases_some}
\end{figure}

\begin{figure}[!htb]
 \centering
\includegraphics[width=11cm]{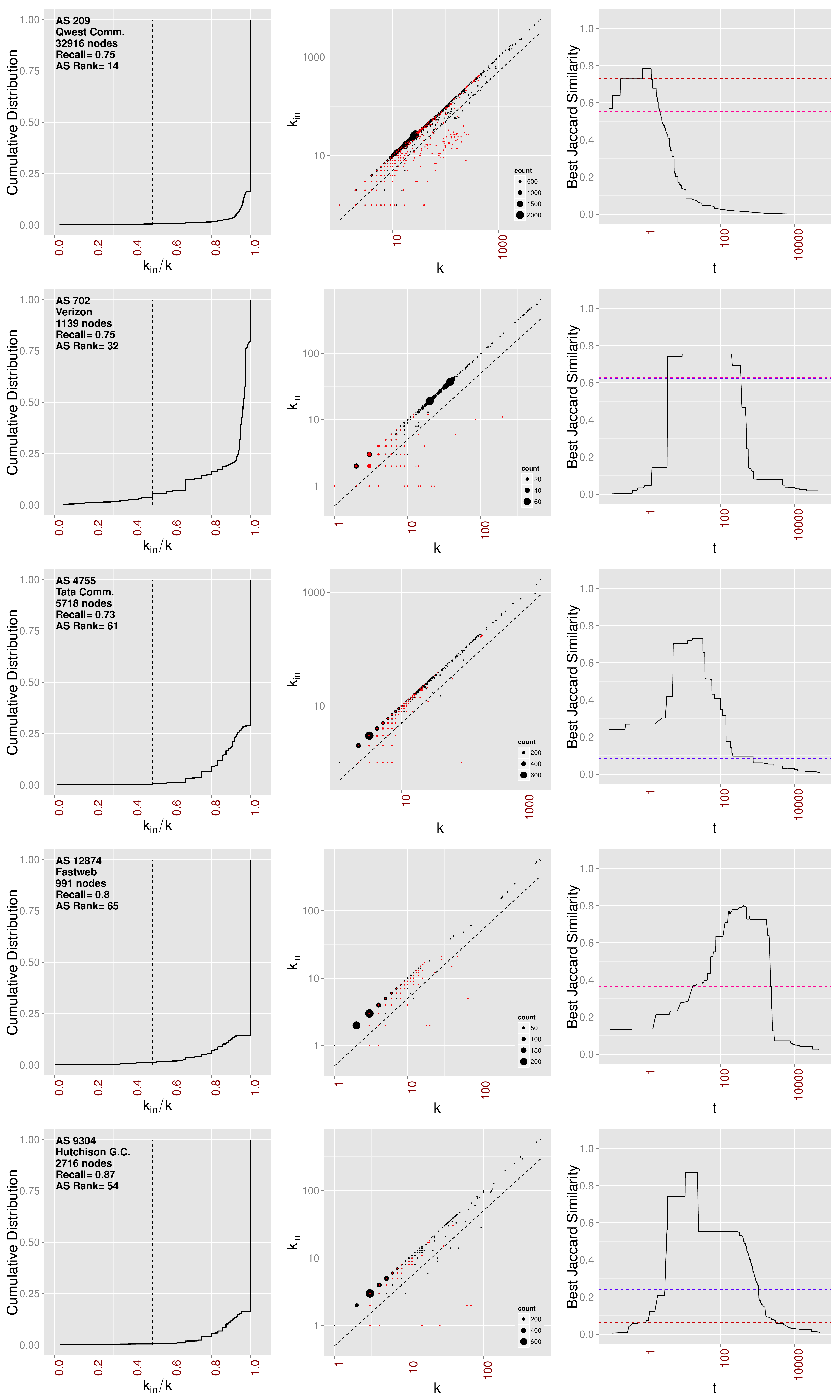}
  \caption{{High-ranked ASes with good recall score.} Normalized internal degree distribution {\em (left-column)}, $k_{in}$ vs $k$ {\em (middle-column)} and the best Jaccard similarity at resolution $t$ {\em (right-column)} of the Multiresolution Deltacom algorithm for a set of 5 ASes with at least $100$ nodes and high AS Rank, retrieved with a good recall score. A more detailed description can be found in Figure~\ref{ases_some}.}\label{ases_some2}
\end{figure}

\begin{figure}[!htb]
 \centering
\includegraphics[width=11cm]{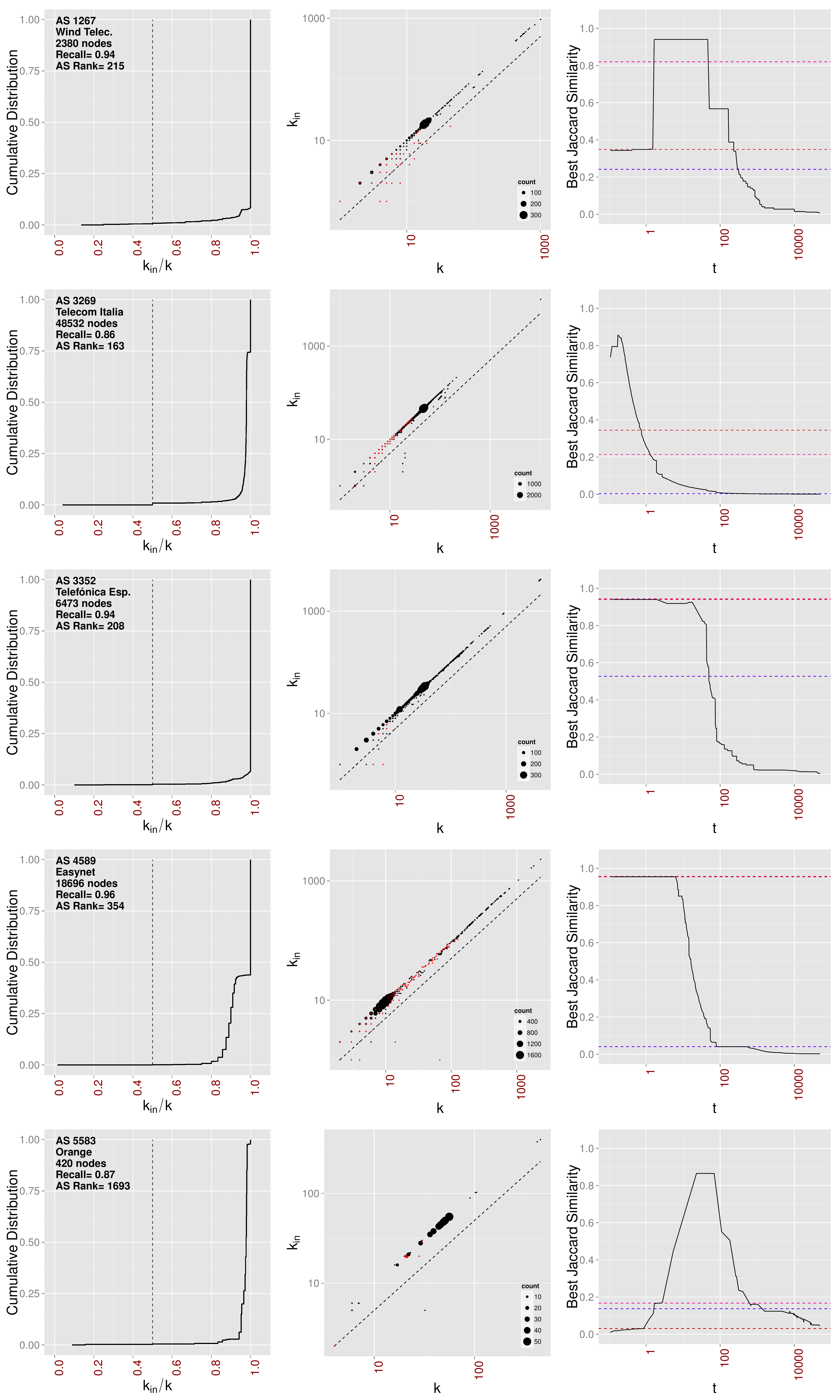}
  \caption{{Average-ranked ASes with very good recall score.} Normalized internal degree distribution {\em (left-column)}, $k_{in}$ vs $k$ {\em (middle-column)} and the best Jaccard similarity at resolution $t$ {\em (right-column)} for a set of 5 ASes with at least $100$ nodes and intermediate AS Rank, retrieved with a very good recall score. A more detailed description can be found in Figure~\ref{ases_some}.}\label{ases_some3}
\end{figure}

In Figures~\ref{ases_some}, \ref{ases_some2} and~\ref{ases_some3} we analyse some particular cases of {\em (a) well-detected, high-ranked ASes}, {\em (b) bad-detected, high-ranked ASes} and {\em (c) very well-detected, mean-ranked ASes}. Each row represents an AS, and the three columns represent {\em (i) the internal node degree distribution} (normalized by the node degree), {\em (ii) the internal node degree vs. node degree} and {\em (iii) the evolution of the best Jaccard similarity at different resolutions of the Multiresolution Deltacom algorithm} (its maximum corresponds to the AS recall score).

In the central pictures of Figure~\ref{ases_some} we plot the number of internal edges as a function of node degree for some ASes which were not well-detected. The diagonal dashed line corresponds to $k_{in}/k$; the nodes under this line have more external connections than internal ones. The colour indicates if the node was classified inside the community which retrieved the best Jaccard similarity {\em (black)} or not {\em (red)}. We observe that all the ASes in Figure~\ref{ases_some} have high degree nodes with more external that internal connections; these nodes are not recognized as part of their ASes and thus induce many of the short-degree nodes into error. Into these group of ASes we found carrier networks and IXCs (e.g., AT\&T and Sprint), IXPs (e.g., the London Internet Exchange) and some CDNs (e.g., Akamai and Amazon).

These types of ASes lay in the backbone of the Internet and they are densely connected among them, so that their nodes may not respect the classical notions of community. It is possible that we should think in an overlapping community structure for the core of the Internet, following the ideas in~\cite{lesko_2015} regarding the core-periphery structure of networks and its relation with community structure. In this network, highly overlapping communities might be related with the presence of IXPs or nodes which do not satisfy the strong community definition, for example.

\begin{figure}[!htb]
 \centering
 \includegraphics[width=12cm]{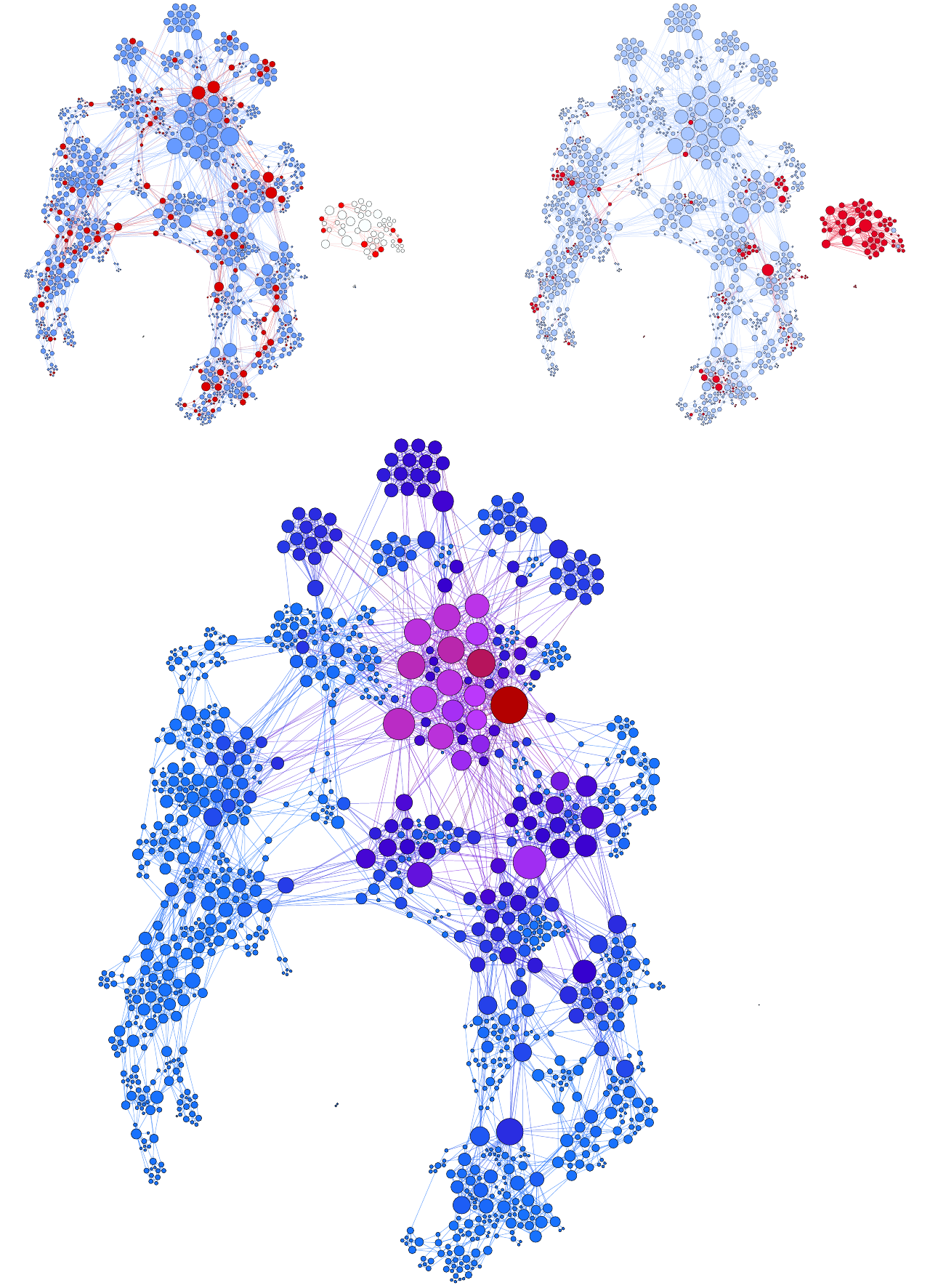}
  \caption{{The AS 15169 (Google).} In the upper left picture, Google's AS 15169 with a sample of $15\%$ of the nodes coloured in red. The gray nodes correspond to $\hat{C}_{AS}^{(2)}$, the community found by the Multiresolution Deltacom algorithm using that sample and the AS size. Lower, the detected community, $\hat{C}_{AS}^{(2)}$. In the upper right picture, all the AS with the frontier nodes coloured in red. The recall score for this AS is $R_3=0.93$. (Image generated with Gephi)}\label{fig_google}
\end{figure}

\begin{figure}[!htb]
 \centering
 \includegraphics[width=12cm]{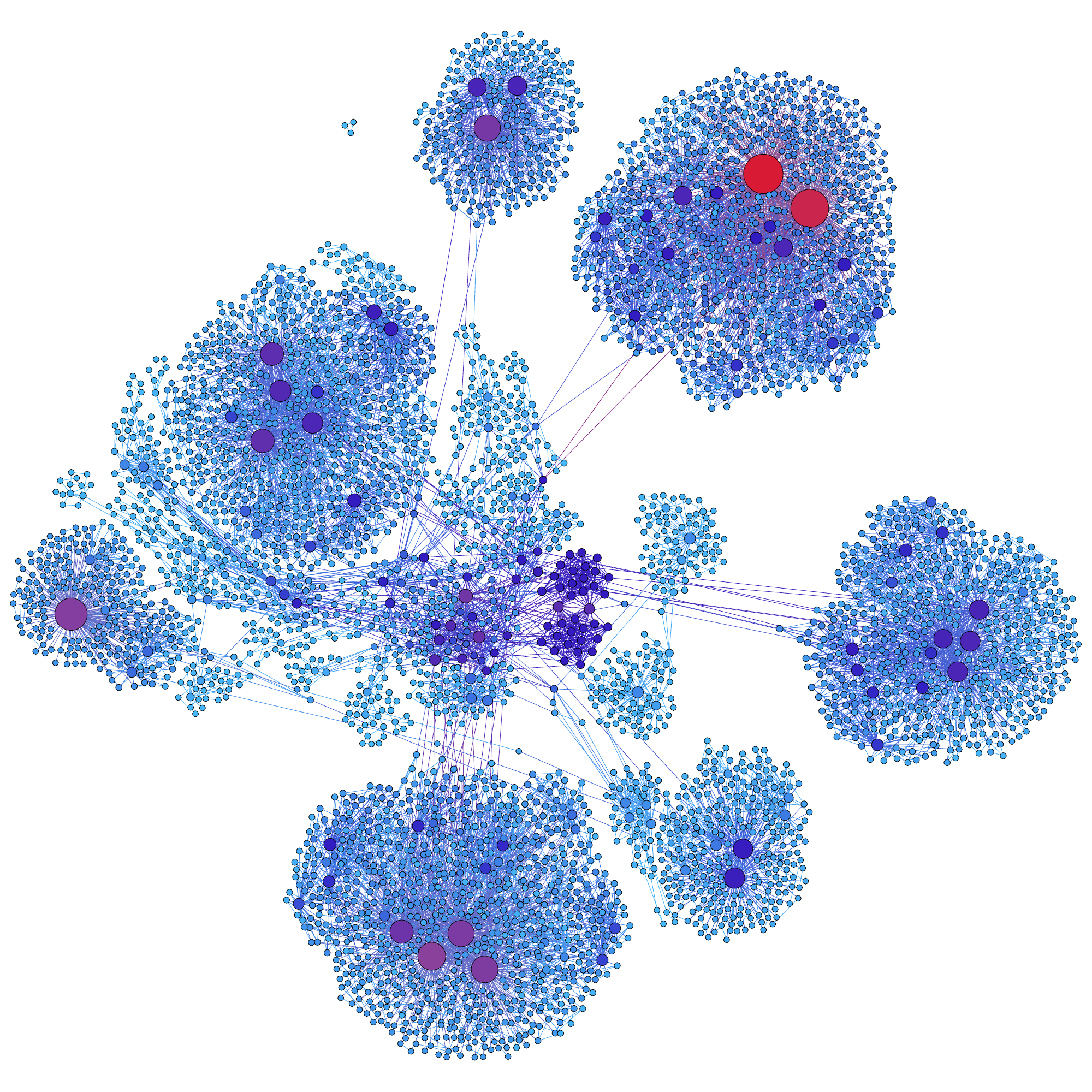}
  \caption{{The AS 17506 (Ucom Corporation, Japan).} The structure of this Japanese AS with $4739$ nodes is organized into different clusters. Many of these clusters have a redundancy structure formed by $4$ routers, which is a typical configuration in many ASes. Though the clusters can be clearly distinguished, they also constitute one large community which the Multiresolution Deltacom algorithm retrieves with a recall score of $R_2=0.78$. Some nodes play a special role into this cohesion by acting as hubs. We can identify them by computing the eigenvector centrality, which is shown as the colour of the nodes in a light blue/blue/violet/red scale: the red nodes have the highest centrality values, and they are the main hubs. The classical modularity optimization throws a recall of $0.19$ instead, because the AS is already combined with others at $t=1$. On the other hand, Infomap obtains a recall of $0.13$. (Image generated with Gephi)}\label{fig_ucom}
\end{figure}

ASes with good recall score are shown in Figures~\ref{ases_some2} and~\ref{ases_some3}. In them, all the high-degree nodes have a high proportion of internal connections. These nodes usually act as hubs for their ASes; into this category we find medium-sized and large-sized ISPs, university networks and other small ASes. An example of one of these networks is the Google AS, shown in Figure~\ref{fig_google}. In the lower part of the figure we plot the AS 15169 as detected by the Multiresolution Deltacom algorithm using the $15\%$ of the nodes and the AS size; the recall score of the AS is $0.93$. The colour of the nodes represents their eigenvector centrality inside the community with a light blue/blue/violet/red scale for increasing centrality values; the radius represents their degree in the full graph. The upper left picture shows the $15\%$ sample in red, and the upper right figure shows the nodes in the frontier of the AS, also in red. Those nodes from the AS which were not detected by our algorithm are shown in white in the upper left picture. We observe an interesting modular structure, with several small modules strongly connected among them, and with a tree-like structure. Some high-degree nodes act as hubs inside the AS; however, they are not connected to other ASes usually. The AS has also a small connected component, probably due to a bias in the exploration or to the limitations of the AS affiliation algorithm; this small component was not detected. However, the reconstruction of the main component was excellent; the only difference being some small degree nodes not belonging to the AS, but to carrier ASes like AT\&T and Level3. Almost all the nodes in the frontier of the main component have small degree and small centrality values. The role of hubs is also visualized in Figure~\ref{fig_ucom} for the UCom Corporation AS.

It is interesting to contrast the general results against some classical notions of community. Looking at the central pictures of Figures~\ref{ases_some}, \ref{ases_some2} and~\ref{ases_some3} again, we can see that all of them have nodes under the dashed diagonal line which represents the limit of the notion of strong community by Radicchi {\em et al.}~\cite{radicchi2004}, i.e., having more internal connections than external ones. This notion is quite strict indeed, and neither the real ASes nor the communities found by modularity maximization follow it. In~\cite{hu2008comparative}, Hu {\em et al.} introduced a relaxed version of this notion, in which each node has more internal connections than towards each of the other communities separately; we found that $4\%$ of the nodes (spread through $70\%$ of the ASes) do not follow this definition, either.

Hric {\em et al.}~\cite{PhysRevE.90.062805} have recently used the Jaccard similarity for identifying communities with ground-truth groups as we did. They also observed that community detection algorithms perform very poorly according to this measure and they proposed some conjectures. One of these is the need to change the notion of community. Our work may shed some light into this question: in the Internet router-level graph, we show that the coexistence of communities with multiple resolutions is the main difficulty for the detection algorithms. So, when proposing new conceptions of community structure, we should take into account that very heterogeneous communities must coexist. 

Finally, we state that we exploited the multiresolution ability of our Deltacom algorithm for stopping at every possible resolution. It might be possible to apply this type of technique into other methods, in some way. For example, the CommUGP algorithm is a growing process which traverses the graph finding one community after another. It has a very strict cut criterion aimed at finding very cohesive structure; however, the Autonomous Systems might appear as consecutive nodes during the process. Regarding the Infomap algorithm, we observed that its performance was much better than that of classical modularity or LPM, but it does not overcome the resolution limit problem. In fact, Kawamoto and Rosvall~\cite{kawamoto} showed in a recent paper that the resolution limit also affects Infomap's map equation, though it is less restrictive than in the modularity functional.

\clearpage

\section{Conclusions}

We analysed the internal community structure of the Internet Autonomous Systems by using Deltacom, a multiresolution algorithm. Deltacom performs community detection through modularity optimization at different resolution levels in one single run, and with a low computational cost. The algorithm was made available to the scientific community as an open-source software through SourceForge.

We applied Deltacom to the router-level graph of the Internet obtained from CAIDA, and we observed that most of the ASes which are not in the Internet backbone can be identified as communities if the proper resolution level is used in the community discovery process. However, we showed that classical modularity optimization failed to detect these Autonomous Systems due to the large variety of resolutions that they have.

Many of the Autonomous Systems in the Internet backbone (and usually identified by a very high AS rank) could not be identified as communities, and we observed that this was due to the presence of high degree nodes with more external connections than internal ones, violating one of the main notions of community structure. Instead, stub ASes and ISP's have high-degree nodes which act as hubs and provide internal connectivity, while the external connectivity is usually provided by lower-degree nodes. These ASes were well detected as communities.

Finally, by correlating the resolution of each AS with its size and using a small fraction of the AS affiliation data, we showed that most of the ASes which do have a community structure can be retrieved. This provides a method for identifying Autonomous Systems with scarce information: a short sample of their nodes and their size.

We think that these results on the Internet structure at the router level can be used for improving Internet topology models and the estimation of properties as node centrality or clustering, and shall also lead to new discussions regarding the hierarchical structure of the Internet.

\section*{Acknowledgements}	
  The authors thank Jorge R. Busch for his suggestions and discussions on the Deltacom algorithm. This work was funded by an UBACyT 2012-2015 grant (20020110200181) and a PICT-Bicentenario 2010-01108. M.~G.~B. also acknowledges a CONICET fellowship.

\bibliographystyle{plain} % Style BST file
\bibliography{epj_data_science-IR}      % Bibliography file (usually '*.bib' )

\end{document}